\documentclass[aps,prd,nofootinbib,showpacs,superscriptaddress, preprint]{revtex4-1}
\pdfoutput=1
\usepackage[utf8]{inputenc}
\usepackage{amsmath,amssymb}
\usepackage{epsfig}
\usepackage{comment}
\usepackage{graphicx}
\usepackage[usenames,dvipsnames]{color}
\usepackage{float}
\usepackage{bbold}
\usepackage[colorlinks,citecolor=blue]{hyperref}
\usepackage{color}
\usepackage{subfigure}

\begin{document}
\title{Affleck-Dine Cogenesis of Baryon and Dark Matter}


\author{Debasish Borah}
\email{dborah@iitg.ac.in}
\affiliation{Department of Physics, Indian Institute of Technology Guwahati, Assam 781039, India}

\author{Suruj Jyoti Das}
\email{suruj@iitg.ac.in}
\affiliation{Department of Physics, Indian Institute of Technology Guwahati, Assam 781039, India}

\author{Nobuchika Okada}
\email{okadan@ua.edu}
\affiliation{Department of Physics, University of Alabama, Tuscaloosa, Alabama 35487, USA}

\begin{abstract}
We propose a mechanism for cogenesis of baryon and dark matter (DM) in the universe via the Affleck-Dine (AD) route. An AD field which breaks the lepton number symmetry, leads to the generation of lepton asymmetry by virtue of its cosmic evolution, which then gets transferred into lepton and dark sectors. While the lepton asymmetry gets converted into baryon asymmetry via sphalerons, the dark sector asymmetry leads to the final DM abundance with the symmetric part being annihilated away due to resonantly enhanced annihilation, which we choose to be provided by a gauged $B-L$ portal. Stringent constraints from DM direct detection forces DM and $B-L$ gauge boson masses to be light, in the few GeV ballpark. While a large portion of the model parameter space is already ruled out, the remaining parameter space is within sensitivity of laboratory as well as cosmology based experiments. The AD field also plays the role of inflaton with the required dynamics by virtue of its non-minimal coupling to gravity, consistent with observations.
\end{abstract}

\maketitle

\section{Introduction}
The origin of the observed baryon asymmetry and dark matter (DM) in the universe \cite{Aghanim:2018eyx, Zyla:2020zbs} have been longstanding puzzles in particle physics. While DM comprises approximately $27\%$ of the present universe, the highly asymmetric baryonic matter part gives rise to around $5\%$ of total energy density. Since the standard model (SM) of particle physics can not explain these two observed phenomena, several beyond standard model (BSM) proposals have been put forward. Among different BSM scenarios for particle DM, the weakly interacting massive particle (WIMP) has been the most widely studied one \cite{Kolb:1990vq, Jungman:1995df, Bertone:2004pz, Feng:2010gw, Arcadi:2017kky, Roszkowski:2017nbc}. On the other hand, the mechanism of baryogenesis \cite{Weinberg:1979bt, Kolb:1979qa} which invokes out-of-equilibrium decay of heavy new particles, has been the most popular scenario to explain the baryon asymmetry of the universe (BAU). One appealing way to achieve baryogenesis, while connecting it to neutrino physics, is the leptogenesis \cite{Fukugita:1986hr} route where a non-zero lepton asymmetry is first generated which later gets converted into the BAU via electroweak sphalerons \cite{Kuzmin:1985mm}.

Although these well-motivated frameworks can explain BAU and DM independently, the very similarity between their abundances namely, $\Omega_{\rm DM} \approx 5\,\Omega_{\rm Baryon}$ gives rise to another puzzle deserving an explanation. Ignoring any numerical coincidence or anthropic origin, similar baryon-DM abundance can be generated simply by finding a common generation or cogenesis mechanism\footnote{See \cite{Boucenna:2013wba} for a mini review on such cogenesis mechanisms.}. Such cogenesis mechanisms can be broadly categorised into asymmetric dark matter (ADM) \cite{Nussinov:1985xr, Davoudiasl:2012uw, Petraki:2013wwa, Zurek:2013wia, Barman:2021ost, Cui:2020dly} where DM sector also has an asymmetry and WIMPy baryogenesis where BAU is generated from DM annihilations \cite{Yoshimura:1978ex, Barr:1979wb, Baldes:2014gca, Chu:2021qwk, Cui:2011ab, Bernal:2012gv, Bernal:2013bga, Kumar:2013uca, Racker:2014uga, Dasgupta:2016odo, Borah:2018uci, Borah:2019epq, Dasgupta:2019lha, Mahanta:2022gsi}. Usually in ADM scenarios, out-of-equilibrium decay of a field to dark and visible sectors is responsible for creating similar asymmetries. Here, we consider the Affleck-Dine (AD) mechanism \cite{Affleck:1984fy} to be the common origin of dark and visible sector asymmetries. Contrary to earlier works in this direction \cite{Cheung:2011if, vonHarling:2012yn} where supersymmetric scenarios of Affleck-Dine baryogenesis and DM were discussed, we consider a simple non-supersymmetric setup where baryogenesis occurs via leptogenesis. In a typical AD mechanism of this type, a lepton number (L) carrying field $\Phi$, to be referred to as the AD field hereafter, breaks L explicitly by virtue of its quadratic term. The cosmological evolution of the AD field then leads to the generation of lepton asymmetry. We show that this asymmetry can get transferred to lepton and dark sectors leading to the required cogenesis. Considering the AD field to be an SM singlet scalar, we adopt a Dirac neutrino scenario such that $\Phi$ field transfers the L asymmetry into a right handed neutrino ($\nu_R$) first which subsequently gets transferred to lepton doublets. The same $\Phi$ field also couples to DM, a Dirac fermion singlet, in order to transfer the asymmetry to dark sector. The symmetric part of DM annihilates via additional gauge interactions, provided by a gauged $B-L$ setup. The stringent constraints from DM direct detection \cite{LZ:2022ufs} forces DM to lie in few GeV ballpark with similar mass window for $B-L$ gauge boson within reach of several future experiments. Due to Dirac nature of light neutrinos with $B-L$ gauge interactions, the parameter space of the model can be tightly constrained from cosmic microwave background (CMB) experiments like Planck via measurement of effective relativistic degrees of freedom $N_{\rm eff}$ while the remaining parameter space within reach of next generation CMB experiments. In addition to all these, the AD field can also play the role of inflaton which, by virtue of its self quartic coupling and non-minimal coupling to gravity can give rise to the required inflationary parameters, as constrained by CMB data \cite{Akrami:2018odb, BICEP:2021xfz}.

This paper is organised as follows. In section \ref{sec1} we briefly outline the model, followed by the discussion of dynamics of the AD field, predictions for inflationary parameters as well as lepton asymmetry in section \ref{sec2}. In section \ref{sec3} we summarise our results incorporating different experimental constraints along with future sensitivities and finally conclude in section \ref{sec4}.

\section{The model}
\label{sec1}
As mentioned earlier, we consider a gauged $B-L$ extension of the SM \cite{Davidson:1978pm, Mohapatra:1980qe, Marshak:1979fm, Masiero:1982fi, Mohapatra:1982xz, Buchmuller:1991ce} to realise AD cogenesis. This not only ensures the inclusion of right handed neutrinos due to anomaly cancellation requirements, but also provides a portal for DM annihilations. While the $B-L$ charges of the SM fields are straightforward, the newly introduced fields and their quantum numbers are shown in table \ref{tab:Lepto1}. Two scalar singlet fields $\Phi, \Phi'$ with non-zero $B-L$ charges are chosen such that the former plays the role of the AD field while the latter takes part in spontaneous symmetry breaking. Since the $B-L$ symmetry is broken by a scalar field with $B-L$ charge $-4$, it prevents the generation of Majorana mass of $\nu_R$, a requirement for light Dirac neutrino scenario. A Dirac fermion $\chi$, singlet under the SM gauge symmetry, is introduced to play the role of DM while a neutrinophilic Higgs doublet $H_2$ plays the role of generating light Dirac neutrino mass of sub-eV scale. Additional discrete symmetries $Z_2 \times Z^D_2$ are introduced in order to prevent unwanted couplings. Out of these, softly broken $Z_2$ leads to generation of light neutrino mass while unbroken $Z^D_2$ keeps DM stable.

\begin{table}
    \centering
    \begin{tabular}{|c|c|c|c|}
    \hline
    Fields & $SU(3)_c \times SU(2)_L \times U(1)_Y$  & $U(1)_{B-L}$ & $Z_2 \times Z^D_2$\\
    \hline  
   $ \nu_{R}$ & $(1,1,0)$ & $-1$ & $(-1, 1)$ \\
      $ \chi_{L,R}$ & $(1,1,0)$ & $-1$ & $(1,-1)$ \\
        $H_2$ & $(1,2,-1/2)$ & $0$ & $(-1, 1)$ \\
          $ \Phi$ & $(1,1,0)$ & $2$ & $(1, 1)$ \\
          $ \Phi'$ & $(1,1,0)$ & $-4$ & $(1, 1)$ \\
          \hline
    \end{tabular}
    \caption{BSM field content of the model.}
  \label{tab:Lepto1}
\end{table}

 The relevant part of the model Lagrangian is 
\begin{align}
    \mathcal{L} \supset \mathcal{L}_{\rm SM} + \mathcal{L}_{\rm inf} (\Phi,R) - Y_\nu \overline{L} \tilde{H_2} \nu_R -  M_\chi \overline{\chi}\chi - Y_R \overline{\nu^c_R} \nu_R \Phi-Y_D \overline{\chi^c} \chi \Phi - V(\Phi, \Phi') +{\rm h.c.},
\end{align}
where
\begin{align}
   V(\Phi, \Phi')= m_{\Phi}^2 |\Phi|^2  + \lambda_{\Phi} |\Phi|^4 + \mu \Phi' \Phi^2 + \lambda' \left(|\Phi'|^2-\frac{v_{BL}^2}{2}\right)^2 - \lambda_{\rm mix} |\Phi|^2 |\Phi'|^2 + {\rm h.c.}
\end{align}
Here $v_{BL}$ is the vacuum expectation value (VEV) of the singlet scalar $\Phi'$ responsible for breaking gauged $B-L$ symmetry. In the above Lagrangian, $\mathcal{L}_{\rm inf} (\Phi,R)$ denotes the non-minimal coupling of the inflaton / AD field to gravity, which is of the form $\mathcal{L}_{\rm inf} (\Phi,R)=-\frac{1}{2} \left( M_P^2 + \xi |\Phi|^2 \right)R$, and is required to provide successful inflationary predictions. While other scalar fields can also couple non-minimally to gravity, we consider the corresponding couplings to be negligible.

Neutrinos remain Dirac in such a setup with light Dirac neutrino mass arising from a tiny VEV of $H_2$ \cite{Davidson:2009ha}. The soft $Z_2$ breaking term $\mu^2_{\rm soft} H_2 H^\dagger_1$ with $H_1$ being the SM Higgs doublet, leads to an induced VEV of the neutral component of $H_2$, thereby generating a light Dirac neutrino mass. Once $\Phi'$ acquires a non-zero VEV, it generates $Z'$ mass \cite{Nath:2021uqb} which can be light enough and plays role in annihilating out the symmetric part of asymmetric DM $\chi$.

\section{Dynamics of the AD field}
\label{sec2}
\begin{figure}[t]
$$
\includegraphics[scale=0.6]{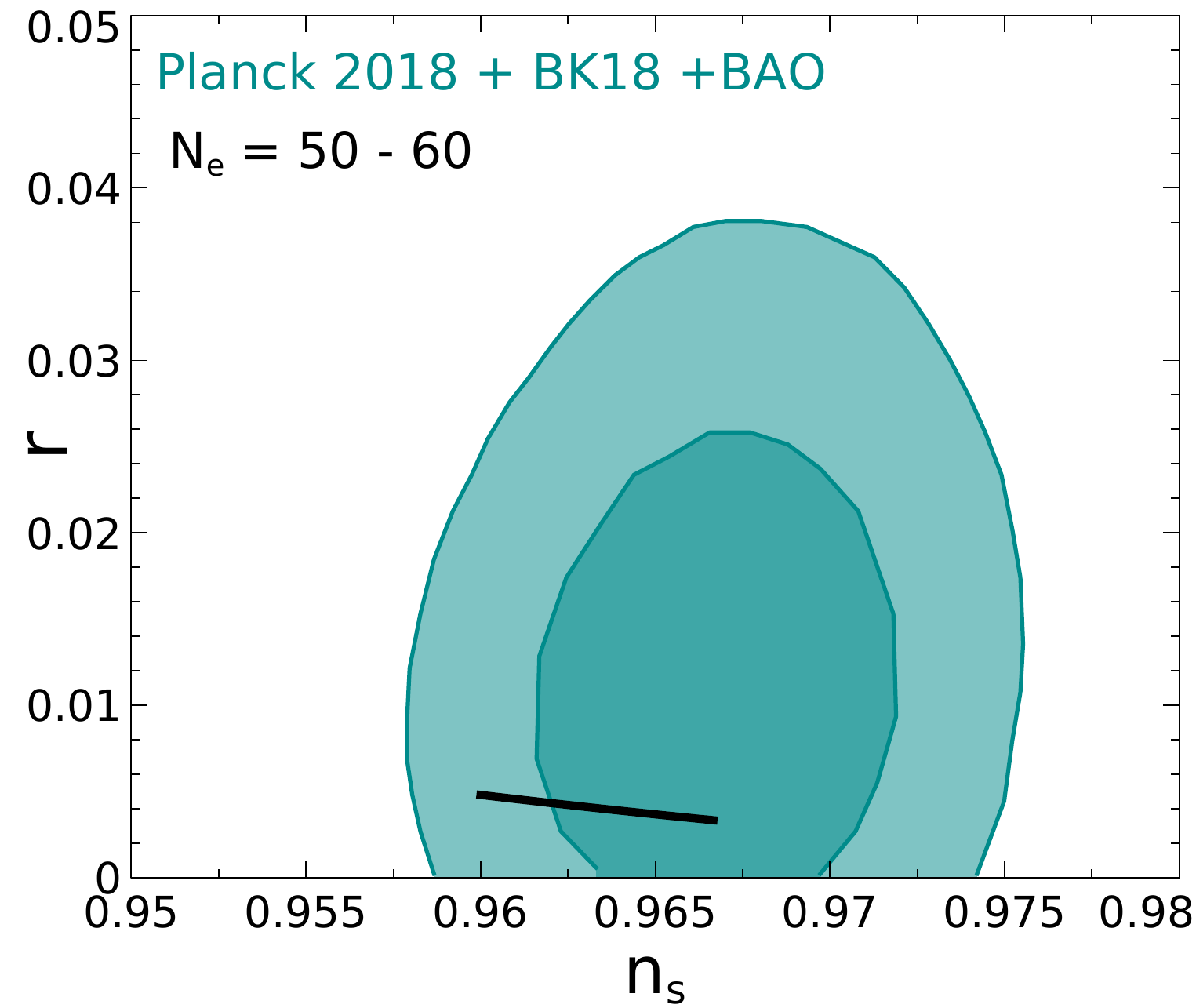}
$$
\caption{$n_{s}-r$ contour for $\xi \gg 1$, by varying $N_e$ from 50 to 60 (from left to right) along with the Planck 2018 1$\sigma$ and 2$\sigma$ bounds \cite{BICEP:2021xfz}.}
\label{fig:nsr}
\end{figure}

We identify the AD field $\Phi$ to be the inflaton to realise the Higgs inflation via non-minimal coupling to gravity \cite{Bezrukov:2007ep}. Earlier works on a common origin of inflation and baryogenesis or leptogenesis via AD mechanism can be found in \cite{Charng:2008ke, Hertzberg:2013jba, Hertzberg:2013mba, Takeda:2014eoa, Babichev:2018sia, Cline:2019fxx, Cline:2020mdt, Lin:2020lmr, Lloyd-Stubbs:2020sed, Kawasaki:2020xyf, Mohapatra:2021aig, Barrie:2021mwi, Mohapatra:2022ngo}. The non-minimal coupling of $\Phi$ is $\frac{\xi}{2} \phi^2 R$ where $R$ represents the Ricci scalar and $\xi$ is a dimensionless coupling of the singlet scalar to gravity. When $\Phi> M_P/\sqrt{\xi}$, it slow-rolls and causes inflation by virtue of its non-minimal coupling to gravity. Such a scenario provides tensor-to-scalar ratio and scalar spectral index \cite{Okada:2010jf, Okada:2015lia, Borah:2020wyc, Borah:2021inn}, which are consistent with cosmological data from CMB experiments like Planck \cite{Akrami:2018odb} and BICEP/Keck \cite{BICEP:2021xfz}. For example, with $\xi\gg 1$, we have predictions for inflationary observables, namely the magnitude of spectral index ($n_s$) and tensor-to-scalar ratio ($r$) as $r=0.003$, $n_{s}=0.967$ for number of e-folds $N_e=60$, which satisfies Planck 2018 data at 1$\sigma$ level \cite{Akrami:2018odb}. For completeness, in Fig. \ref{fig:nsr}, we show the inflationary predictions  in the $n_{s}-r$ plane varying $N_e$, along with the most recent Planck + BICEP/Keck bounds \cite{BICEP:2021xfz}. 


Now, during inflation, an effective negative mass squared term  $- \lambda_{\rm mix} \langle|\Phi|^2\rangle$ is generated for $\Phi'$. Assuming $\lambda' v_{BL}^2 \ll \lambda_{\rm mix} \langle |\Phi|^2 \rangle$, $\Phi'$ develops a B-L symmetry breaking VEV $\gg$ $v_{BL}$. As $|\Phi|$ goes down after inflation, $\langle\Phi'\rangle$ becomes smaller and eventually arrives at $\langle\Phi'\rangle=v_{BL}/\sqrt{2}$. This generates the lepton number violating term $A\Phi^2= \epsilon m_{\Phi}^2 \Phi^2$ from $\mu \Phi' \Phi^2$ term in the scalar potential (where $\epsilon =\frac{\mu v_{BL}}{\sqrt{2} m_{\Phi}^2}$), which is the key to generate asymmetry via the AD mechanism. This should happen before $\Phi$ reaches $\Phi^*= \frac{m_{\Phi}}{\sqrt{\lambda_{\Phi}}}$, which represents the moment when the quadratic oscillation of $\Phi$ begins. Hence, we require $\lambda' v_{BL}^2 \gg \lambda_{\rm mix} \langle |\Phi^*| \rangle^2 \sim \frac{\lambda_{\rm mix}}{\lambda_{\Phi}}m_{\Phi}^2$.   

\subsection{Cogenesis of baryon and DM}

\begin{figure}[htb!]
$$
\includegraphics[scale=0.5]{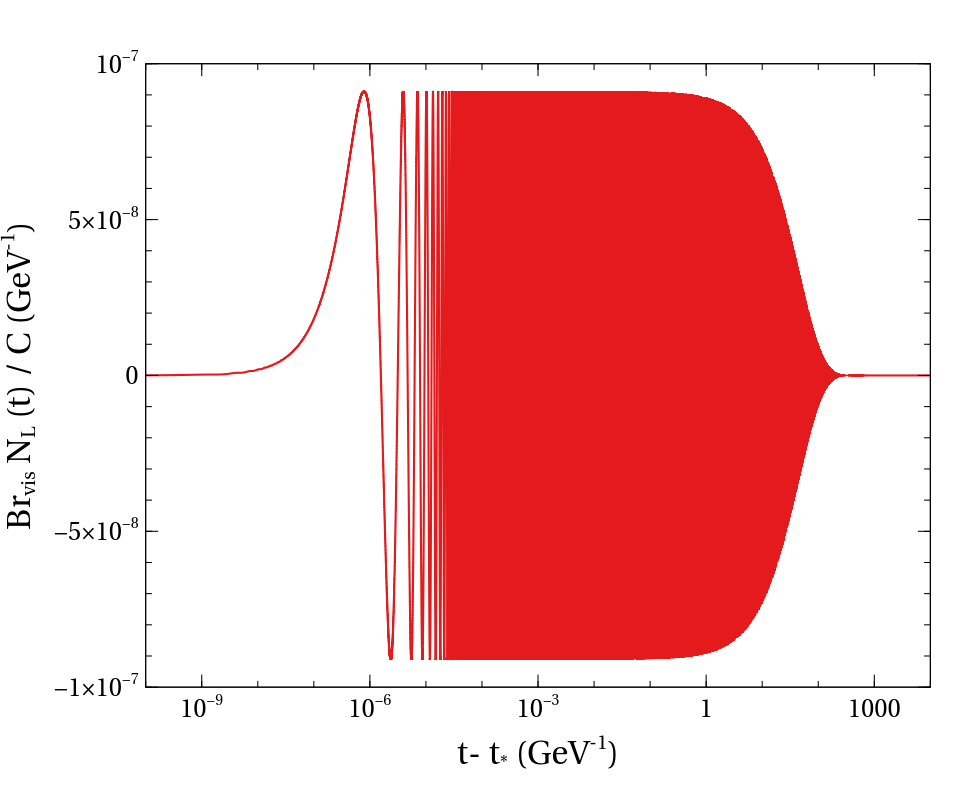}~~~~
\includegraphics[scale=0.5]{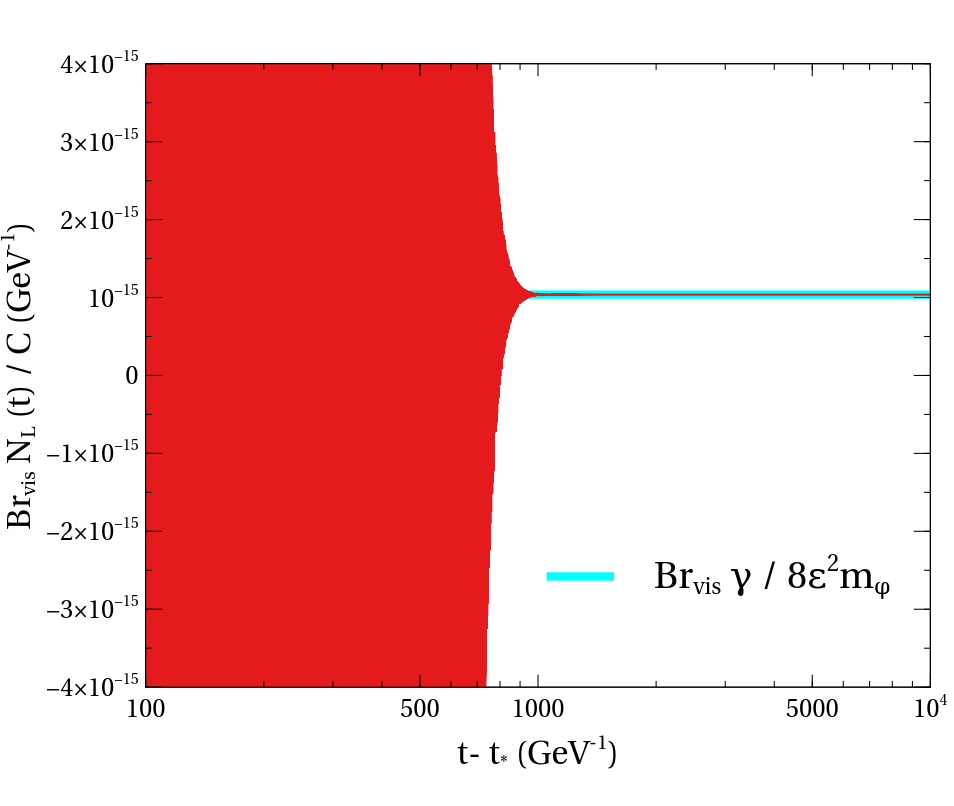}
$$
$$
\includegraphics[scale=0.5]{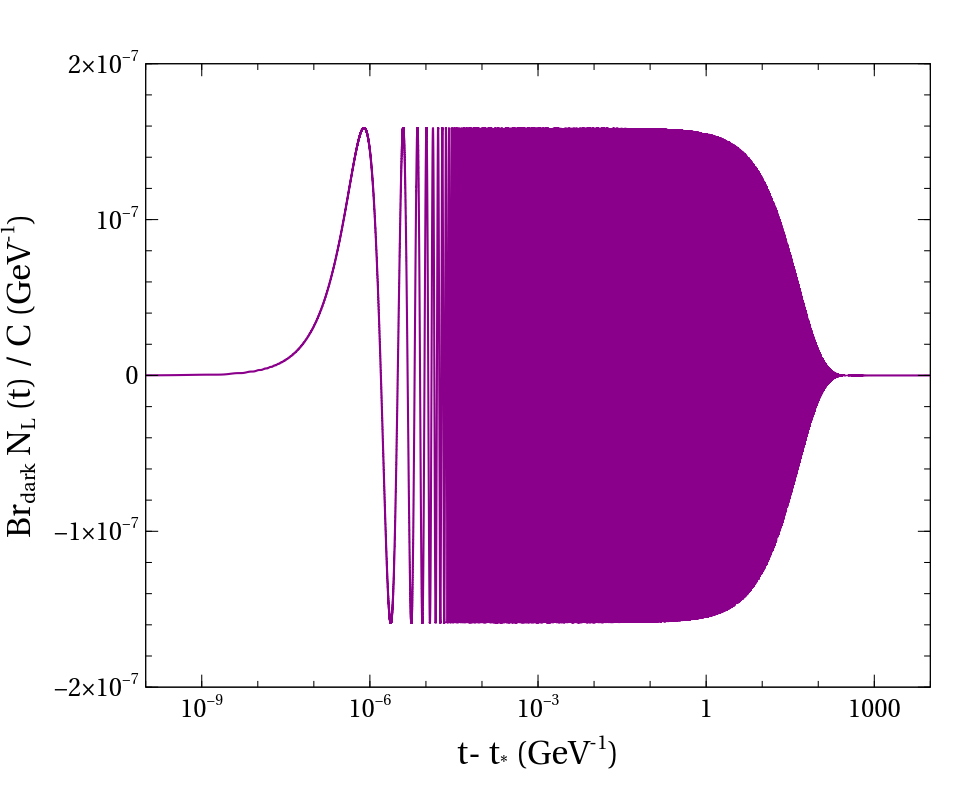}~~~~
\includegraphics[scale=0.5]{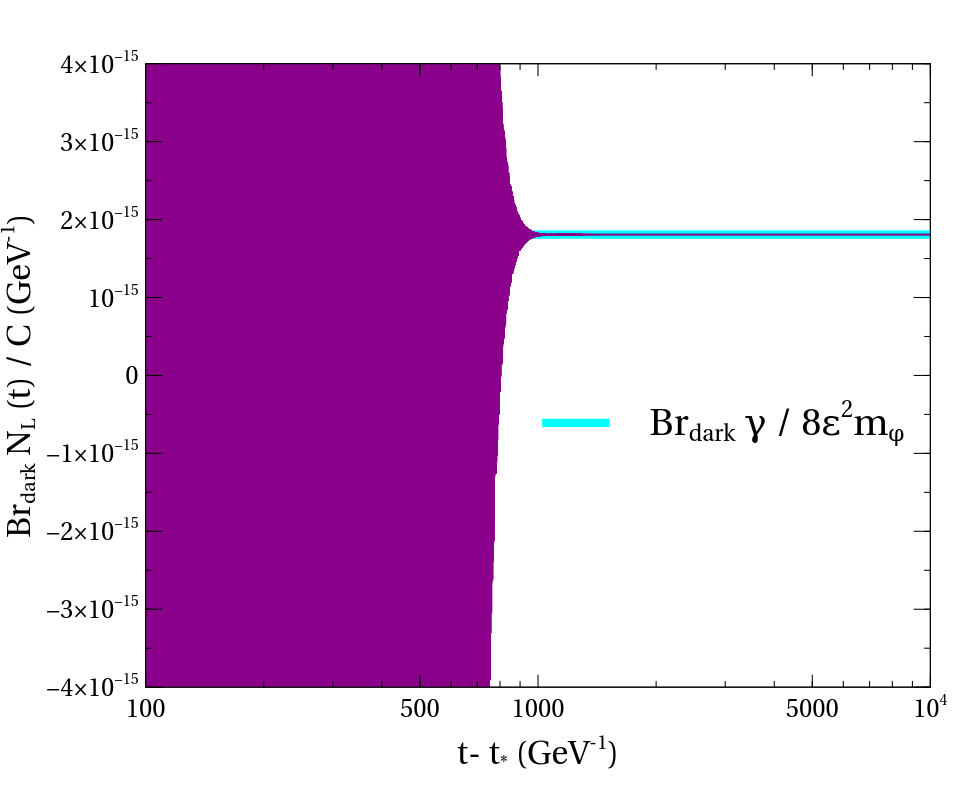}
$$
\caption{Evolution of the comoving asymmetry (normalised by the factor C) and its transfer to the visible sector (top panel) and dark sector (bottom panel), for the BP set 1 of Table \ref{tab:BP}. The right panel shows the evolution near $t\sim 1/\Gamma_{\Phi}$, where the oscillation amplitude exponentially decays to reach the constant asymptotic value (Eq. \eqref{eqn:NB}) shown by the cyan line.}
\label{fig:asym-evol}
\end{figure}

For $ \Phi^*=\frac{m_{\Phi}}{\sqrt{\lambda_{\Phi}}} \lesssim\Phi\lesssim \frac{M_{P}}{\sqrt{\xi}}$, quartic term $\lambda_{\Phi} \lvert \Phi \rvert^{4}$ dominates and $\Phi\propto1/a$. Once $\Phi$ reaches $\Phi^*$, difference in the real and imaginary values of $\Phi$ creates an asymmetry in the $\Phi$ condensate which oscillates with period $T_{\rm asy}=\frac{\pi}{\epsilon m_{\Phi}}$. Now, the comoving asymmetry generated for $t>t^{*}$ can be written as \cite{Mohapatra:2021aig} 
\begin{align}
    N_L (t)&\simeq 4 Q_\Phi \, A \, \phi_{1,I} \, \phi_{2,I} \left(\frac{\phi_I}{\Phi^*} \right) 
 \int_{t_*}^t dt' \, {\cos}(m_{1}(t'-t_*)) \, {\cos}(m_{2}(t'-t_*))  \, e^{- \Gamma_\Phi (t'-t_*)} 
\end{align}
where $\Gamma_\Phi$ indicates the total decay rate of the inflaton $\Phi$ to $\nu_{R}$ and $\chi$. $\phi_{1,I} \, \phi_{2,I}$ indicate the initial values of the real and imaginary parts of $\Phi$ and $\phi_I=\sqrt{(\phi_{1,I})^2+(\phi_{2,I})^2}$, whereas $m^2_1=m^2_\Phi-2A$ and $m^2_2=m^2_\Phi +2A$. Asymmetry created is  transferred to visible and dark sectors through decays $\ensuremath{\Phi\rightarrow\nu_{R}}\nu_{R}$
and $\ensuremath{\Phi\rightarrow\chi\chi}$ respectively, which also reheats the Universe with reheat temperature $T_{R}\simeq\sqrt{\Gamma_{\Phi}M_{P}}$ \footnote{While the inflaton field $\Phi$ also has quartic Higgs portal interactions like $\Phi^\dagger \Phi H^\dagger_i H_i$ with $i= 1,2$, we consider such quartic couplings to be sufficiently small in order to keep the explosive production of Higgs bosons via parametric resonance sub-dominant \cite{Kofman:1994rk, Kofman:1997yn, Greene:1997fu}. This validates our estimate of the reheat temperature from perturbative decay of inflaton.}. In our scenario $2 A \gg \Gamma_\Phi m_\Phi$ and in this limit, for $t\gtrsim 1/\Gamma_{\phi}$, the integral above reduces to the constant value 
\begin{align}
    N_L (t)&\simeq C \frac{\gamma}{8 \epsilon^2 m_{\Phi}},
    \label{eqn:NB}
\end{align}
where $\gamma=\Gamma_{\Phi}/m_{\Phi}$ and $C= 4 Q_\Phi \, A \, \phi_{1,I} \, \phi_{2,I} \left(\frac{\phi_I}{\Phi^*} \right)$. In Fig. \ref{fig:asym-evol}, we show the evolution of the comoving asymmetry $N_{L} (t)$, which gets transferred to the visible sector (top panel) and dark sector (bottom panel), depending on the branching ratio of the inflaton decay, ${\rm Br}_{\rm vis}$ and ${\rm Br}_{\rm dark}$,  respectively. The asymmetry initially rises from zero and then oscillates until $t\gtrsim 1/\Gamma_{\Phi}$, when its amplitude exponentially damps to reach the constant value given by equation \eqref{eqn:NB} multiplied by a factor of ${\rm Br}_{\rm vis, dark}$.

The final asymmetry generated in visible and dark sectors can be written as  \cite{Lloyd-Stubbs:2020sed,Mohapatra:2021aig}
\begin{equation}
\frac{n_{L}}{s}\simeq Q_{\Phi} {\rm Br}_{\rm vis}\frac{T_{R}^{3}}{\epsilon m_{\Phi}^{2}M_{P}}\simeq10^{-10},\,\,\,\,\,\label{eq:vis}
\end{equation}
\begin{equation}
\frac{n_{\rm DM}}{s}\simeq Q_{\Phi} {\rm Br}_{\rm dark}\frac{T_{R}^{3}}{\epsilon m_{\Phi}^{2}M_{P}}\simeq\frac{0.12}{2.75\times10^{8}}\left(\frac{M_{{\rm DM}}}{\rm GeV}\right)^{-1},\label{eq:dark}
\end{equation}
where Br denotes the branching ratios of $\Phi$ decay. The above equations hold under the conditions $\epsilon\ll1$ and $\epsilon m_{\Phi}/\Gamma_{\Phi}\gg 1$, which is satisfied in our model.  The reheat temperature which enters into the above equations must be less than $m_{\Phi}$, for the asymmetry to survive. Writing $T_{R}= K m_{\Phi}$ with $K<1$, Eq. \eqref{eq:vis} and \eqref{eq:dark} give 
\begin{equation}
m_{\Phi}\simeq\frac{10^{-10}\epsilon}{Q_{\phi} {\rm Br}_{\rm vis}K^{3}}M_{P},\label{eq:mphi1}
\end{equation}
\begin{equation}
m_{\Phi}\simeq\frac{0.12}{2.75\times10^{8}}\frac{\epsilon}{Q_{\Phi} {\rm Br}_{\rm dark}K^{3}}\left(\frac{M_{{\rm DM}}}{\rm GeV}\right)^{-1}M_P.\label{eq:mphi2}
\end{equation}
Equating the above equations gives the following relation for the DM mass, which is determined solely by the branching ratios of $\Phi$ decay 
\begin{equation}
M_{\rm DM}=4.36\left(\frac{{\rm Br}_{\rm vis}}{{\rm Br}_{\rm dark}}\right) \text{GeV}.\label{eq:mdm}
\end{equation}

The asymmetry in $\nu_R$ can be transferred to the lepton doublets via Yukawa interactions with neutrinophilic Higgs $H_2$ through lepton-number conserving processes such as $\nu_{R} \nu_{R} \leftrightarrow L L $  proportional to $Y_{\nu}^4$  or $\nu_{R} H_2 \leftrightarrow L\, Z/W $ proportional to $Y_{\nu}^2$. Such processes should be thermalised before the electroweak scale $T_{\rm EW}$ such that the lepton asymmetry stored in lepton doublets gets converted into baryon asymmetry via sphalerons. Considering the process whose rate depends upon $Y_{\nu}^4$, this leads to a condition $T_{\rm EW}^{3}\frac{Y_{\nu}^{4}}{T_{\rm EW}^2}\gtrsim \sqrt{\frac{\pi^2}{90}g_*} \frac{T_{\rm EW}^{2}}{M_{P}}$. This can be realized through Yukawa couplings $Y_{\nu}\gtrsim 10^{-4}$. Considering the process which goes as $Y_{\nu}^2$, it leads to a weaker bound $Y_{\nu}\gtrsim 10^{-6}$. This gives an upper bound on the $H_{2}$ VEV $v_2$ around keV, since the lightest neutrino mass $m_\nu\sim Y_{\nu} v_2 = \mathcal{O}(0.1\, {\rm eV})$.

Now, note that in our scenario, the presence of lepton number violating interaction given by $\epsilon$ can lead to the washout of the generated asymmetry. This can happen through scatterings with  $\Delta L=2$ :  $\nu_R \nu_R \leftrightarrow \chi \chi$ or $\Delta L=4$ : $\nu_R \nu_R \leftrightarrow \overline{\nu}_R~\overline{\nu}_R$, mediated by $\Phi$ exchange and the $\epsilon$ term. If the decoupling temperature of such process is higher than the reheat temperature $T_R$, the washout effect would be absent. Thus, the following condition must hold (considering the $\Delta L=4$ washout)
\begin{equation}
T_{R}^{3}\frac{Y_{R}^{4}\epsilon^2T_R^2}{4 \pi m_{\Phi}^{4}}\lesssim \sqrt{\frac{\pi^2}{90}g_*} \frac{T_{R}^{2}}{M_{P}}.\label{eq:washout}
\end{equation}

It should be noted that a more minimal scenario for AD cogenesis, in principle, is possible if we consider DM to be the neutral component of a vector like fermion doublet. In such a case, a single decay process of the AD field produces lepton and dark sector asymmetries. Since dark and visible sector asymmetries are identical in such a setup, it forces the DM mass to be in few GeV ballpark. Since DM is part of an electroweak doublet, direct search constraints rule out such a scenario. This leaves us with the choice discussed above, where dark and visible sector asymmetries depend upon the respective branching ratios of the AD field.

\section{Results and Discussion}
\label{sec3}
After showing the key aspects of AD field evolution and its role in inflation as well as generation of dark and visible sector asymmetries in previous section, we now summarise our results in terms of relevant model parameters in view of existing experimental constraints as well as future sensitivities. \\

\noindent
{\it Annihilation of the symmetric DM:} Since in our scenario, the relic abundance of DM is determined by the asymmetric component only, it is important to make sure that the symmetric part of DM abundance annihilates away. This can be guaranteed by the $B-L$ interactions mediated by the gauge boson $Z'$. In order to enhance the cross-section such that the symmetric DM abundance is negligible, we consider DM mass near the $Z'$ resonance, $M_{\rm DM}\simeq M_{Z'}/2$. While it is possible to have a sufficiently large DM annihilation cross-section, the same $B-L$ portal interactions of DM also lead to a large spin-independent DM-nucleon cross-section, tightly constrained by direct search experiments \cite{LZ:2022ufs}. In order to evade these stringent bounds, we consider lighter DM mass around a few GeV. In order to maintain resonantly enhanced annihilation condition, this also requires the $B-L$ gauge boson to be light. Using the narrow width approximation, the condition for observed DM relic density for the symmetric part can be written as \cite{Nath:2021uqb} 
\begin{equation}
    g_{\rm BL}\simeq 3.46 \times 10^{-5} \frac{M_{Z'}}{10 ~ \rm GeV}.
    \label{eqn:gbl}
\end{equation}
Since the cross-section in this narrow width approximation varies as $\langle \sigma v \rangle \propto g_{\rm BL}^{2}$, the symmetric DM relic density goes as $\Omega_{\rm DM} h^{2}|_{\rm sym}\propto 1/g_{\rm BL}^{2}$. Hence, increasing $g_{\rm BL}$ by a factor of say, 10 from the one given by Eq. \eqref{eqn:gbl}, decreases the relic density to $\Omega_{\rm DM} h^{2}|_{\rm sym}=0.01 \, \Omega_{\rm DM} h^{2}|_{\rm sym}^{\rm observed}$. This ensures that the major portion of DM relic is the asymmetric part generated together with baryon asymmetry via AD mechanism. \\

\noindent
{\it Predictions for $\Delta {\rm N_{eff}}$:}
An interesting aspect of such light Dirac neutrino scenarios is the enhancement of the effective relativistic degrees of freedom $N_{\rm eff}$ which can be probed at CMB experiments, as can be found in recent works \cite{Abazajian:2019oqj, FileviezPerez:2019cyn, Nanda:2019nqy, Han:2020oet, Luo:2020sho, Borah:2020boy, Luo:2020fdt, Mahanta:2021plx, Biswas:2021kio, Borah:2022obi, Okada:2022cby, Borah:2022enh}. The current $2\sigma$ limit (95$\%$ CL) on $N_{\rm eff}$ from the Planck 2018 data is $N_{\rm eff}= 2.99^{+0.34}_{-0.33}$ \cite{Aghanim:2018eyx}, consistent with the SM prediction $N^{\rm SM}_{\rm eff}=3.045$. Future CMB experiment CMB Stage IV (CMB-S4) is expected reach a much better sensitivity of $\Delta {\rm N}_{\rm eff}={\rm N}_{\rm eff}-{\rm N}^{\rm SM}_{\rm eff}
= 0.06$ \cite{Abazajian:2019eic}. Assuming all three $\nu_R$ to get thermalised in the early universe and decouple instantaneously above the electroweak scale, simple entropy conservation arguments lead to $\Delta N_{\rm eff} \approx 0.14$ \cite{Abazajian:2019oqj}. If we consider the 2$\sigma$ limit from Planck 2018 data $\Delta \rm N_{eff}$ $\lesssim0.28$, it requires $\nu_R$ decoupling temperature $T_D$ just above QCD phase transition, i.e. $T_D \gtrsim$ 300 MeV. Using 
\begin{equation}
    n_{\nu_{R}} \langle \sigma v \rangle |_{T= T_D} = H (T_D)
\end{equation}
where $n_{\nu_{R}}$ denotes the equilibrium number density and $\langle \sigma v \rangle$ is the thermal averaged cross section of $\nu_{R}$ \cite{Okada:2022cby}, we get an upper bound on the gauge coupling $g_{\rm BL}$ in $M_{Z'} \gg T_D$ limit as 
\begin{equation}
    g_{\rm BL}\lesssim g_{\rm BL}^{\rm max} = 1.17 \times 10^{-4} \frac{M_{Z'}}{\rm GeV}\label{eq:neff}
\end{equation}
in order to be in agreement with Planck 2018 bounds. However, for $M_{Z'} \sim T_D$, the interaction rate of $\nu_R$ can get resonantly enhanced, leading to much stronger limit on $g_{\rm BL}$ \cite{Heeck:2014zfa}. Although $\nu_R$ can get thermalised via Yukawa interactions too, we consider gauged $B-L$ portal to be more dominant due to light $Z'$.

\begin{figure}[t]
$$
\includegraphics[scale=0.6]{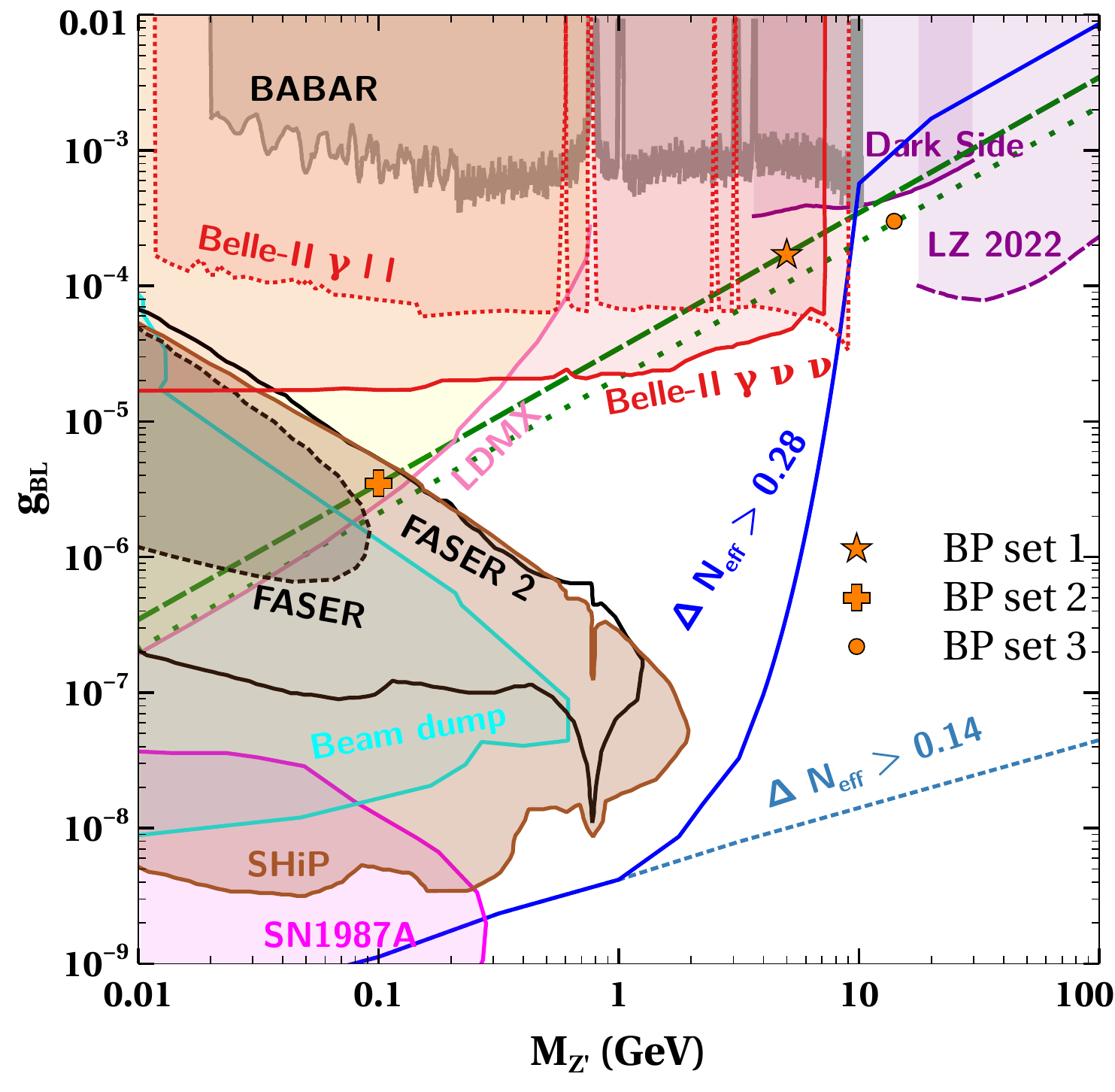}
$$
\caption{Summary plot in  $g_{\rm BL}-M_{Z'}$ plane with $M_{\rm DM}\simeq M_{Z'}/2$ showing relevant constraints and future sensitivities. The green dashed (dotted) contour corresponds to the parameter space for which the symmetric part of DM gives $\sim 1 \%$ ($\sim 3\%$) of the observed DM relic density. The region above the solid blue contour is excluded from Planck 2018 constraints on $\Delta N_{\rm eff}$. The three benchmark points given in Table \ref{tab:BP} are also highlighted.}
\label{fig:b-l}
\end{figure}

\begin{table}[t]
\begin{center}
\begin{tabular}{|c||c||c|c|}\hline
Parameter & BP (set 1) & BP (set 2)& BP (set 3)\\\hline
$\epsilon$ & $10^{-3}$ & $10^{-5}$ & $4.4 \times 10^{-3}$\\
$K$ & $0.1$ & $0.2$ & $0.095$\\
$m_\Phi$& $10^9$ GeV& $1.3\times 10^7$ GeV & $10^9$ GeV\\
$M_{\rm DM}$ & $2.5$ GeV & $0.05$ GeV & $7$ GeV\\
$M_{Z'}$ & $5$ GeV & $0.1$ GeV & $14$ GeV\\
$Y_R$ & $10^{-5}$ & $3.16\times10^{-7}$ & $10^{-5}$\\
$Y_D$ & $1.32 \times 10^{-5}$ & $2.95 \times 10^{-6}$ & $7.89 \times 10^{-6}$\\
$g_{\rm BL}$ & $1.7\times 10^{-4}$ & $3.5 \times 10^{-6}$ & $3 \times 10^{-4}$\\\hline
\end{tabular}
\end{center}
\caption{Three sets of benchmark parameters highlighted in Fig. \ref{fig:b-l}.}
\label{tab:BP}
\end{table}

  In Fig. \ref{fig:b-l}, we summarise our results in $g_{\rm BL}-M_{Z'}$ plane assuming $M_{\rm DM}\simeq M_{Z'}/2$. We show the $\chi \, \bar{\chi}$ annihilation rate required to keep the symmetric DM abundance at $\sim 1 \%$ ($\sim 3 \%$) of the observed thermal DM relic density by green dashed (dotted) contour such that the symmetric component gives a sub-dominant contribution compared to the asymmetric component generated by AD mechanism. For points above this green contours, the symmetric DM abundance will be further suppressed. The three points marked on these contours correspond to the set of model parameters given in table \ref{tab:BP}. The benchmark parameters shown in this table are consistent with DM relic requirements but face stringent experimental constraints. For example, the Planck 2018 $2\sigma$ bound on $\Delta {\rm N_{eff}}$, shown by the solid blue contour, disfavours a major portion of this plane. The bound in light $Z'$ regime is so strong that it rules out two of the benchmark points given in table \ref{tab:BP} which correspond to symmetric DM abundance at $\sim 1 \%$ of the observed DM relic. In fact, all points on the green dashed contour corresponding to this relative abundance of symmetric DM component are ruled out either by $\Delta {\rm N_{ eff}}$ bound or DM direct detection bounds from LZ 2022 \cite{LZ:2022ufs} and DarkSide-50 \cite{DarkSide:2018bpj}. The third benchmark point is currently allowed from all such constraints and correspond to $3\%$ contribution of symmetric DM component to total relic. It is to be noted that almost the entire plane shown in Fig. \ref{fig:b-l} can be probed by future measurements of $\Delta {\rm N_{ eff}}$ at  SPT-3G \cite{SPT-3G:2019sok} and CMB-S4 \cite{CMB-S4:2016ple} experiments. The excluded regions from dark photon search by electron-positron collider BABAR \cite{BaBar:2014zli}, various beam dump experiments \cite{Riordan:1987aw, Bjorken:1988as, Bross:1989mp, Konaka:1986cb, Davier:1989wz,  Banerjee:2019hmi, Blumlein:1990ay, Blumlein:2013cua, Bergsma:1985qz, Astier:2001ck, Bernardi:1985ny}, supernova SN1987A observation \cite{Kamiokande-II:1987idp,Hong:2020bxo,Shin:2021bvz} are also shown in Fig. \ref{fig:b-l}, along with the DM direct detection bound from LZ 2022 \cite{LZ:2022ufs} and DarkSide-50\cite{DarkSide:2018bpj}. The expected sensitivity of various upcoming experiments including Belle-II \cite{Dolan:2017osp}, FASER \cite{Feng:2017vli,FASER:2018eoc,FASER:2019aik}, LDMX \cite{Berlin:2018bsc}, SHiP \cite{Alekhin:2015byh} are also shown. Upcoming experiment DUNE will also be able to probe some part of the currently allowed parameter space as has been studied in \cite{Asai:2022zxw, Chakraborty:2021apc}. We, however, do not show the corresponding sensitivity curve in Fig. \ref{fig:b-l} for simplicity.

  As seen from the above discussion, the scenario discussed here is very predictive and hence tightly constrained, allowing only a tiny parameter space consistent with asymmetric dark matter relic and baryon asymmetry of the universe. Clearly, the constraints from Planck 2018 bounds on $\Delta {\rm N_{eff}}$ is the most stringent one due to the Dirac nature of light neutrinos in our setup. It is possible to have other realisations of AD cogenesis where light neutrinos can be of Majorana nature and hence $\Delta {\rm N_{eff}}$ bounds do not arise. One possibility is to consider inverse seesaw realisation of light neutrino masses \cite{Mohapatra:2021ozu} with $\nu_R, H_2$ in our model replaced by $N_R, S_L$ with $U(1)_{B-L}$ quantum numbers $-1, 0$ respectively. The AD field can transfer asymmetry into $N_R$ which can then be converted into lepton doublets via Yukawa interactions of type $Y_\nu \overline{L} \tilde{H_1} N_R$ with $H_1$ being the SM Higgs doublet. The same AD field also transfers a part of the asymmetry into DM $\chi$ realising the asymmetric dark matter scenario. The lepton number violating term responsible for inverse seesaw can be generated by another singlet scalar, which can be made to acquire a VEV only at temperatures below the sphaleron decoupling without introducing any new washout processes to affect the asymmetries \cite{Mohapatra:2021ozu}. Thus, the final results remain same as what we have discussed for our model but with more available parameter space due to the absence of $\Delta {\rm N_{eff}}$ bounds.

  It should be noted that, in the minimal model, we have relied upon resonant annihilation of the symmetric DM component in order to be consistent with direct detection bounds. This has led to a very precise relation between DM and $Z'$ mass $M_{\rm DM}=\frac{M_{Z'}}{2}(1-\delta)$ where $\delta$ is a tiny number \cite{Nath:2021uqb}. The fine-tuning in $\delta$ can be reduced by relaxing the upper bound on symmetric DM component. While a large symmetric DM abundance is undesirable from cogenesis point of view, it is possible to get rid of such fine-tuned resonance enhancement in non-minimal scenarios. For example, existence of additional light singlet scalars can open up efficient DM annihilation channels which do not require resonance enhancement.

\section{Conclusion}
\label{sec4}
We have proposed a baryon-DM cogenesis scenario via the Affleck-Dine mechanism. The Affleck-Dine field not only leads to the generation of asymmetries in visible sector and dark sectors, but also plays the role of the inflaton field with successful inflationary dynamics by virtue of its non-minimal coupling to gravity. The visible sector asymmetry is generated in the lepton sector first which gets converted into baryon asymmetry via sphalerons. This also leads to a connection to the origin of light neutrino masses, another observed phenomena not explained by the standard model. We choose a gauged $B-L$ portal for DM to satisfy the requirement of annihilating away the symmetric part of DM. The same gauged symmetry also enforces the inclusion of right handed neutrinos, which couple to SM lepton doublets via a neutrinophilic Higgs $H_2$ realising a light Dirac neutrino scenario. The AD field first transfers the asymmetry to DM sector and $\nu_R$ with the latter getting transferred to lepton doublets due to sizeable Yukawa coupling with $H_2$. The tightest constraint in the high mass regime of this scenario comes from the DM direct detection experiments, due to the requirement of a large annihilation cross-section mediated by $Z'$ to annihilate away the symmetric part. This restricts the DM as well as $Z'$ to lie in the light mass regime, around a few GeV, where a large part of the parameter space is already ruled out by laboratory as well as CMB bounds on $N_{\rm eff}$. While some part of the available parameter space can be probed by planned future laboratory based experiments, the future CMB experiments like CMB-S4 will be able to probe it in its entirety. We have also commented on alternative possibilities where stringent $N_{\rm eff}$ bounds can be avoided by considering a Majorana neutrino setup while keeping other results unaffected.

\acknowledgements
The work of N.O. is supported in part by the United States Department of Energy grant DE-SC 0012447. The authors would like to thank Digesh Raut for providing data for the current and prospective bounds from various experiments shown in Fig. 3. The work of D.B. is supported by SERB, Government of India grant MTR/2022/000575. 


\end{document}